\documentclass[preprint,sort&compress,number]{elsarticle}

\usepackage[USenglish]{babel}
\usepackage{epsfig}
\usepackage{amssymb}
\usepackage{amsmath}
\usepackage{bm}
\usepackage{hyperref}
\usepackage{enumerate}

\let\onlinecite\cite

\begin{document}

\begin{frontmatter}

\title{High-performance functional renormalization group calculations for interacting fermions}

\author[theoPhysRWTH]{J. Lichtenstein\corref{cor}}
\ead{lichtenstein@physik.rwth-aachen.de}
\author[theoPhysRWTH]{D. S\'anchez de la Pe\~na}
\author[JSC]{D. Rohe}
\author[AICES,JSC]{E. Di Napoli}
\author[theoPhysRWTH,JARAFIT]{C. Honerkamp}
\author[theoPhysRWTH,theoPhysKoeln]{S. A. Maier}

\cortext[cor]{Corresponding author}

\address[theoPhysRWTH]{Institute for Theoretical Solid State Physics, RWTH Aachen University,\\ D-52074 Aachen, Germany}
\address[JSC]{J\"ulich Supercomputing Centre, Forschungszentrum J\"ulich GmbH,\\ D-52425 J\"ulich, Germany}
\address[AICES]{Aachen Institute for Advanced Study in Computational Engineering Science (AICES), RWTH Aachen University, D-52056 Aachen, Germany}
\address[JARAFIT]{JARA-FIT, J\"ulich Aachen Research Alliance - Fundamentals of\\ Future Information Technology}
\address[theoPhysKoeln]{Institut f\"ur Theoretische Physik, Universit\"at zu K\"oln, D-50937 Cologne, Germany}

\date{July 29, 2016}

\begin{abstract}
  We derive a novel computational scheme for functional
  Renormalization Group (fRG) calculations for
  interacting fermions on 2D lattices. The scheme is
  based on the exchange parametrization fRG for the two-fermion
  interaction, with additional insertions of truncated partitions
  of unity. These insertions decouple
  the fermionic propagators from the exchange propagators and lead to
  a separation of the underlying equations. We demonstrate that
  this separation is numerically advantageous and may pave the way
  for refined, large-scale computational investigations even in the
  case of complex multiband systems. Furthermore, on the
  basis of speedup data gained from our implementation, it is shown
  that this new variant facilitates efficient calculations on a large
  number of multi-core CPUs.  We apply the scheme to the $t$,$t'$
  Hubbard model on a square lattice to analyze the convergence of the
  results with the bond length of the truncation of the partition
  of unity. In most parameter areas, a fast 
  convergence can be observed. Finally, we compare
  to previous results in order to relate our approach to other fRG
  studies.
\end{abstract}

\begin{keyword}
functional Renormalization Group \sep truncated partition of unity \sep interacting fermions \sep hybrid parallelization \sep Hubbard model
\end{keyword}

\end{frontmatter}

\section{Introduction}
Having passed through a process of refinement and development that has
taken more than two decades, the functional Renormalization Group (fRG)
can be rightfully seen as an unbiased method for investigating
interacting Fermi systems. In the medium term, it is conceivable to
use the fRG methods to investigate existing materials in terms of
their ground state properties with {\em quantitative precision}
regarding energy scales and parameter ranges, as well as to discover
new materials with superior features.  While the development process
is far from being completed, with this paper we contribute in pushing
the fRG method forward into this direction.

Our main focus is to show how one can capture the wavevector
dependence of effective two-fermion vertices in a physically
meaningful and numerically advantageous way.  The fRG---as we use it
in this paper---is based on the Wetterich
equation~\cite{Wetterich1993} that describes the evolution of an
effective action. Since this equation results in a full hierarchy of
differential equations encompassing all orders of the
vertex functions, a truncation is necessary in order to ensure
solvability. As defined in the review paper by Metzner
et al.~\cite{Metzner2012}, we build on the level-2 truncation of the
fRG equation hierarchy which has become an often-used
standard. In addition to that, self-energies are
neglected in the current state of the method development for the 
sake of simplicity.  Moreover, in most cases
further approximations are indispensable, for example, 
a discretization of the Brillouin zone (BZ) using a grid of momentum 
sampling points. Within the first fRG studies on Fermi
systems~\cite{Zanchi1998,Halboth2000,Honerkamp2001a}---those 
addressed the 2D Hubbard model---the BZ was discretized by using the
so-called \emph{Fermi surface patching} scheme, where the vertex
dependences on the radial parts of momenta are neglected.  This scheme
was further extended to multiband models and applied to systems like
iron superconductors~\cite{Wang2009,Thomale2009,Lichtenstein2014} or
single- and multilayer
graphene~\cite{Scherer2012,Kiesel2012,Scherer2012a,Pena2014}.

In terms of a method development, the \emph{exchange parametrization
  fRG}---as introduced in Ref.~\onlinecite{Husemann2009} and used,
for example, in
Refs.~\onlinecite{Husemann2012,Giering2012,Eberlein2010,Eberlein2013,Eberlein2014,Eberlein2015,Maier2013}---can
be seen as the next important step. Within that scheme, the
two-particle coupling function, which generally depends on three
external momenta due to the conservation of total
momentum, is decomposed into three single-channel functions, where
every channel only depends strongly on one momentum. As a consequence,
this parametrization---which can be used for the dependence on
frequencies in a similar way~\cite{Karrasch2008}---softens the
scaling of the number of coupled differential equations with
respect to the momentum grid point number. While in the Fermi surface patching
one has to deal with a third power scaling behavior, in the exchange
parametrization the scaling becomes almost linear.
Although there is some freedom in defining the three channels, usually
they can be interpreted as being responsible for charge, spin and pairing
fluctuations respectively. Hence, the exchange parametrization allows to
describe the potentially complex momentum structure of the effective
interaction in a fashion that is physically easier to
interpret.

Besides the exchange parametrization fRG, the \emph{singular-mode fRG
  (SMFRG)}~\cite{Wang2012} was introduced as another scheme to capture
the momentum dependence. Similar to the exchange parametrization fRG,
the SMFRG scheme distinguishes between three different channels, but
it uses other quantities to represent those and it treats the feedback
between these channels differently.\footnote{In a first step the
  contributions are calculated in a single-channel fashion: only
  intra-channel contributions are evaluated. The inter-channel
  feedback is done in a second step by a projection of the
  single-channel results onto the other two.} In this paper we
  build on the two last-named schemes and take the next step in the
  development of the fRG method. In Sec.~\ref{sec:tufrg} we
present a step-by-step derivation of a new fRG variant that combines
features of both the exchange parametrization fRG and the SMFRG, and
we argue that the new variant is numerically beneficial compared to
the exchange parametrization fRG. In order to point out the major
distinction between the latter and the new scheme, we name our newly
developed variant \emph{truncated unity fRG (TUfRG)}. We comment on the relation of the
TUfRG to the SMFRG in \ref{sec:rel_to_smfrg}.

In order to enlarge the application area of the fRG to more complex
systems, method development has to focus on two parallel research
directions: equation parametrization and parallel
implementation. While it is crucial to develop new meaningful
parametrizations and approximations for the flow equations, it is
equally relevant to explore parallelization and performance strategies
enabling the efficient use of massively parallel computing
architectures. For instance, as shown in Ref.~\onlinecite{Rohe2015},
fRG can profit highly from a sophisticated hybrid parallelization that
can achieve a speedup of several orders of magnitude. In
Sec.~\ref{sec:perform}, we discuss our algorithmic choices, their
parallel implementation and the speedup gained by TUfRG when running
on parallel computing platforms. 

The TUfRG contains an additional approximation compared to the
exchange parametrization fRG---namely the insertion of a truncated
partition of unity. This insertion leads to a simplification of the integrals 
involved, which are computationally the most challenging tasks in the 
fRG calculation. In Sec.~\ref{sec:phase_hubb}, we check the quality of
this approximation by applying the scheme to the well
studied~\cite{LeBlanc2015,Honerkamp2001,Giering2012,Eberlein2015,Yamase2016}
$t$-$t'$ Hubbard model. Furthermore, we provide a more analytic view
on this aspect in \ref{sec:app-smooth-ff}.

\section{The TUfRG equations}\label{sec:tufrg}
Since there are comprehensive descriptions of the fermionic fRG already given in other works (for a recent review, see e.g. Refs.~\onlinecite{Metzner2012,Platt2013}), we will not explain 
the basic concept of this method and the derivation of the hierarchy of flow equations in detail, but briefly mention the important equations to bring the 
reader up to speed with our notation.
Afterwards, we will derive the TUfRG from the fermionic fRG equations. For further classification we explain that the TUfRG equations are strongly related to the ones from 
the SMFRG~\cite{Wang2012} and point out the conceptual differences in \ref{sec:rel_to_smfrg}.
\subsection{fRG Flow equations} \label{sec:su2-inv}
 In the following, we consider an effective action of the form
\begin{equation*}
\Gamma [\bar{\psi},\psi] = \int \! d\xi \, \bar{\psi}(\xi) \, Q (k) \, \psi (\xi) +\Gamma^{(4)} [\bar{\psi},\psi] \, ,
\end{equation*}
 where $\xi=(k,\sigma)$, with $k =(k_0,\mathbf{k})$, denotes a collection of frequency ($k_0$), momentum ($\mathbf{k}$) and spin projection ($\sigma$) quantum numbers.
 As described in Refs.~\onlinecite{Salmhofer2001,Metzner2012}, the two-particle interaction
\begin{align*}
 \Gamma^{(4)} [\bar{\psi},\psi] & = \frac{1}{4} \int \! d \xi_1 \dots d \xi_4 \, f(\xi_1,\xi_2,\xi_3,\xi_4) \\
 & \qquad \times \bar{\psi} (\xi_4) \, \bar{\psi} (\xi_3) \, \psi (\xi_2) \, \psi (\xi_1) 
\end{align*}
 of a charge conserving and SU(2) invariant theory can be parametrized
 with one spin-independent coupling function $ V(k_1,k_2,k_3) $ according to
\begin{align*}
 f (\xi_1,\xi_2,\xi_3,\xi_4) & = 
  \left[ V (k_1,k_2,k_3) \, \delta_{\sigma_1, \sigma_4}
 \delta_{\sigma_2, \sigma_3} - V (k_2,k_1,k_3) \, \delta_{\sigma_1, \sigma_3} \delta_{\sigma_2, \sigma_4} \right] \\
  & \quad \times \delta (k_1 +k_2 -k_3 -k_4) \, ,
\end{align*}
 where the $\delta$-function ensures momentum and energy conservation.

\begin{figure}[t]
\centering
Particle-particle diagram $ \mathcal{T}_\mathrm{pp}$\\ \includegraphics[width=0.23\linewidth]{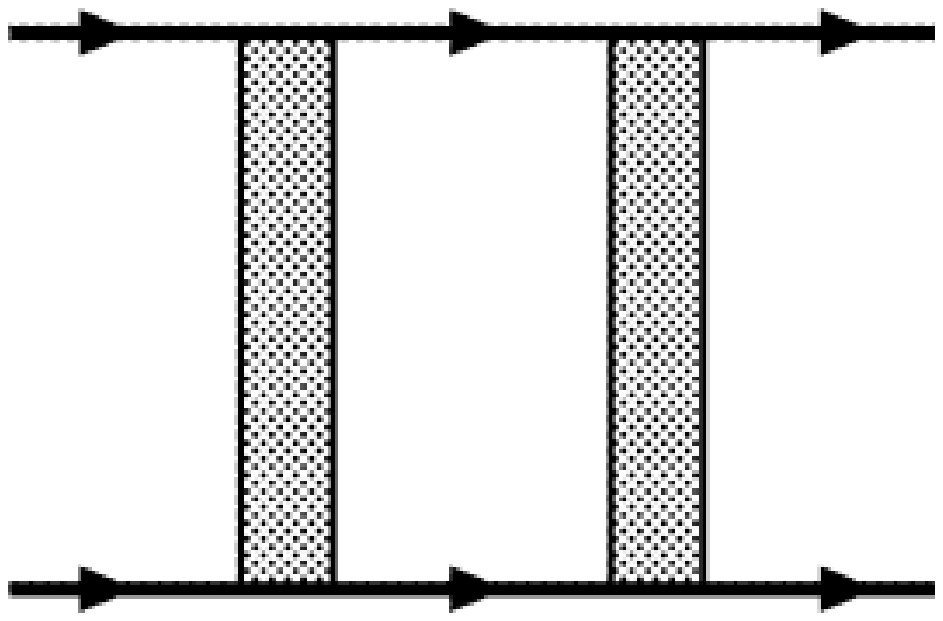} \\[1.5ex]
Crossed particle-hole diagram $ \mathcal{T}_\mathrm{ph}^\mathrm{cr}$\\ \includegraphics[width=0.23\linewidth]{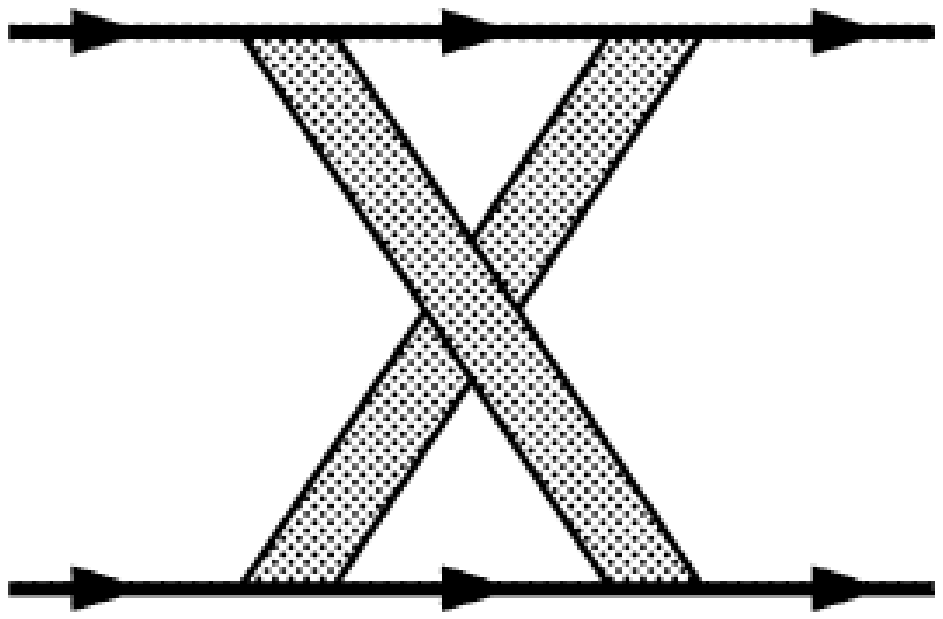} \\[1.5ex]
Direct particle-hole diagrams $ \mathcal{T}_\mathrm{ph}^\mathrm{d}$\\ \includegraphics[width=0.48\linewidth]{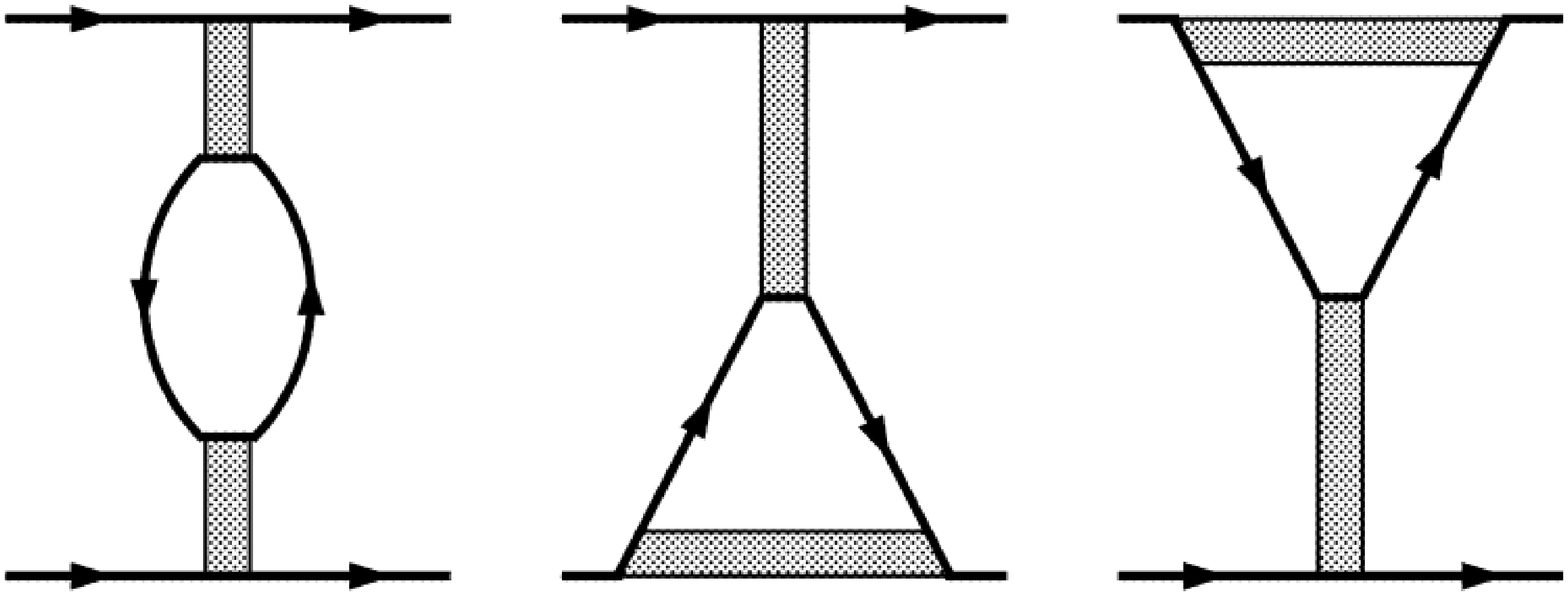}
\caption{The five diagrams driving the flow of the two-particle interaction $ V(k_1,k_2,k_3)$ of an U(1) and $ \mathrm{SU} (2) $ invariant theory. For the closed loops in these diagrams, a scale derivative is implicit. (These figures have been taken from Ref.~\onlinecite{Salmhofer2001}.)} \label{fig:diags-fleq}
\end{figure}

 The quadratic part of the effective action is diagonal in spin-space, which implies 
 \begin{equation*}
  G (\xi_1,\xi_2) = \delta_{\sigma_1,\sigma_2} \delta(k_1-k_2) \, G (k_1) \quad \text{and} \quad
  \Sigma (\xi_1,\xi_2) = \delta_{\sigma_1,\sigma_2} \delta(k_1-k_2) \, \Sigma (k_1)
 \end{equation*}
for the one-particle propagator $G$ and the 1PI self-energy $\Sigma$. Once a regulator is added to the propagator, we can derive the fRG flow equations. 
More explicitly, we replace $G$ by a function $G^\Lambda$ in a way that we get $G^\Lambda\to0$ for $\Lambda\to\infty$ and $G^\Lambda\to G$ for $\Lambda\to 0$. 
This results in differential equations of 1PI vertex functions with respect to the 
regularization scale, which we call $\Lambda$ in this paragraph. The flow equation corresponding to the 1PI self-energy reads
\begin{equation} \label{eqn:sigma-su2fl}
 \dot{\Sigma} (k) = \int\! dp \,\, S(p) \, \left[  V (p,k,p) - 2 V (k,p,p) \right] \, ,
\end{equation}
with the single-scale propagator $S$~\cite{Metzner2012}.
Note that in the following we will not mark dependences on the regularization scale with a superscript $\Lambda$ in order to simplify the notation. A derivative with 
respect to this scale is denoted as a dot above the affected quantity.
The scale derivative of the coupling function $ V $ consists of three parts
\begin{equation*}
 \dot{V} (k_1,k_2,k_3) = \mathcal{T}_\mathrm{pp} (k_1,k_2,k_3)
 +  \mathcal{T}^\mathrm{cr}_\mathrm{ph}  (k_1,k_2,k_3) + \mathcal{T}^\mathrm{d}_\mathrm{ph} (k_1,k_2,k_3) \, .
\end{equation*}
The particle-particle contribution 
\begin{equation} \label{eqn:pp-su2fl}
 \mathcal{T}_\mathrm{pp} = - \int \! d p \, \left[ \partial_\lambda G(p) \, G(k_1+k_2-p) \right]
  V (k_1,k_2,p) \, V(k_1+k_2-p,p,k_3)
\end{equation}
and the crossed particle-hole part 
\begin{equation} \label{eqn:phcr-su2fl}
 \mathcal{T}^\mathrm{cr}_\mathrm{ph} = - \int \! d p \, \left[ \partial_\lambda G(p) \, G(p+k_3-k_1) \right]
 V (k_1,p+k_3-k_1,k_3) \, V(p,k_2,p+k_3-k_1)
\end{equation}
can each be represented by one diagram (see Fig.~\ref{fig:diags-fleq}).
  Vertex corrections and particle-hole screening, however, are accounted for by the direct particle-hole diagrams depicted in Fig.~\ref{fig:diags-fleq}:
\begin{align} \notag
 \mathcal{T}^\mathrm{d}_\mathrm{ph} =  \int \! d p & \,\, \left[ \partial_\lambda G(p) \, G(p+k_2-k_3) \right] 
 \left[ 2 V (k_1,p+k_2-k_3,p) \, V(p,k_2,k_3) \right. \\  \notag
&  -  V (k_1,p+k_2-k_3,k_1+k_2-k_3) \, V(p,k_2,k_3)   \\ \label{eqn:phd-su2fl} & \left.
 -  V (k_1,p+k_2-k_3,p) \, V(p,k_2,p+k_2-k_3) \right] \, .
\end{align}
\subsection{Channel decomposition}
 Let us now recapitulate the channel decomposition of Karrasch \emph{et al.}~\cite{Karrasch2008} for the frequency- and by Husemann and Salmhofer~\cite{Husemann2009} for the momentum-dependence of the vertices.
In these approaches, the coupling function $ V(k_1,k_2,k_3) $ is decomposed into contributions resulting from three different channels. More precisely, we have
\begin{align*}
    V \left(k_1,k_2,k_3 \right)   & = V^{(0)}_{k_1,k_2,k_3}  - \Phi^\mathrm{SC}_{k_1+k_2,\frac{k_1-k_2}{2},\frac{k_4-k_3}{2}}  + \Phi^\mathrm{M}_{k_1-k_3,\frac{k_1+k_3}{2},\frac{k_2+k_4}{2} }  \\ &
 \quad + \frac{1}{2} \Phi^\mathrm{M}_{k_3-k_2,\frac{k_1+k_4}{2},\frac{k_2+k_3}{2} }   - \frac{1}{2} \Phi^\mathrm{K}_{k_3-k_2,\frac{k_1+k_4}{2},\frac{k_2+k_3}{2} }  \, ,
\end{align*}
with $ V^{(0)} $ being the bare interaction, and $ \Phi^{\mathrm{SC}} $, $ \Phi^\mathrm{M} $, and $ \Phi^\mathrm{K} $ representing the coupling functions of the pairing, 
the magnetic, and the forward scattering channel, respectively. The first argument of each channel accounts for the dependence on the total ($\mathbf{k_1}+\mathbf{k_2}$) 
or on one of the transfer momenta ($\mathbf{k_1}-\mathbf{k_3}$ and $\mathbf{k_3}-\mathbf{k_2}$). 
These momentum combinations appear inside the fermionic loops from Fig.~\ref{fig:diags-fleq} and label the most important momentum dependences at weak coupling. 
Regarding the other two (weak) momentum dependences of each channel, we have chosen a convention that is more symmetric than in Refs.~\onlinecite{Wang2012,Wang2013,Xiang2013}.
These single-channel coupling functions are generated during the flow according to the following equations
\begin{align} \label{eqn:flow-PhiP}
 \dot{\Phi}^\mathrm{SC}_{k_1+k_2,\frac{k_1-k_2}{2},\frac{k_4-k_3}{2}} & =  - \mathcal{T}_\mathrm{pp}  \left(k_1,k_2,k_3 \right)  \\ \notag
 \dot{\Phi}^\mathrm{M}_{k_1-k_3,\frac{k_1+k_3}{2},\frac{k_2+k_4}{2}} & =  \mathcal{T}_\mathrm{ph}^\mathrm{cr}  \left(k_1,k_2,k_3 \right)  \\ \notag
 \dot{\Phi}^\mathrm{K}_{k_3-k_2,\frac{k_1+k_4}{2},\frac{k_2+k_3}{2}} & =  - 2 \mathcal{T}_\mathrm{ph}^\mathrm{d}  \left(k_1,k_2,k_3 \right)  + \mathcal{T}_\mathrm{ph}^\mathrm{cr}  \left(k_1,k_2,k_1+k_2-k_3 \right)  \, .
\end{align}
 At the formal level, the channel decomposition may be performed in a different way. Instead of $\Phi^\mathrm{M}$ and $\Phi^\mathrm{K}$, the
 particle-hole channels are accounted for by the coupling functions $\Phi^\mathrm{C}$ and $\Phi^\mathrm{D}$, which flow according to
\begin{align} \label{eqn:flow-PhiC}
 \dot{\Phi}^\mathrm{C}_{k_1-k_3,\frac{k_1+k_3}{2},\frac{k_2+k_4}{2}} & =  \mathcal{T}_\mathrm{ph}^\mathrm{cr}  \left(k_1,k_2,k_3 \right)  \\ \label{eqn:flow-PhiD}
 \dot{\Phi}^\mathrm{D}_{k_3-k_2,\frac{k_1+k_4}{2},\frac{k_2+k_3}{2}} & =  \mathcal{T}_\mathrm{ph}^\mathrm{d}  \left(k_1,k_2,k_3 \right) 
\end{align}
 and enter in the effective interaction as
\begin{align} \notag
 V \left(k_1,k_2,k_3 \right)   & = V^{(0)}_{k_1,k_2,k_3}  - \Phi^\mathrm{SC}_{k_1+k_2,\frac{k_1-k_2}{2},\frac{k_4-k_3}{2}}  + \Phi^\mathrm{C}_{k_1-k_3,\frac{k_1+k_3}{2},\frac{k_2+k_4}{2}} \\ \label{eqn:simple-deco} &
 \quad + \Phi^\mathrm{D}_{k_3-k_2,\frac{k_1+k_4}{2},\frac{k_2+k_3}{2}} \, .
\end{align}
 This latter form of the channel decomposition corresponds to the parametrization of the interaction used in Refs.~\onlinecite{Wang2012,Wang2013,Xiang2013}.
 In the following, we will work with the latter variant, while magnetic and forward scattering channels can be obtained as
\begin{align*}
 \Phi^\mathrm{M}_{k_1-k_3,\frac{k_1+k_3}{2},\frac{k_2+k_4}{2}} &= \Phi^\mathrm{C}_{k_1-k_3,\frac{k_1+k_3}{2},\frac{k_2+k_4}{2}} \\
 \Phi^\mathrm{K}_{k_3-k_2,\frac{k_1+k_4}{2},\frac{k_2+k_3}{2}}  & =  - 2 \Phi^\mathrm{D}_{k_3-k_2,\frac{k_1+k_4}{2},\frac{k_2+k_3}{2}}\\
 & \quad + \Phi^\mathrm{C}_{k_3-k_2,\frac{k_1+k_4}{2},\frac{k_2+k_3}{2}} \, .
\end{align*}

\subsection{Projection onto exchange propagators}

 Let us now describe the dependence of the coupling functions on the weak momenta through a complete set of form factors
 $\{\,f_m (\mathbf{k})\,\}$ that are square integrable on the BZ. 
 For the particle-particle channel, we can project the single-channel coupling function $ \Phi^\mathrm{SC}$ onto a matrix
 $\mathbf{P} (l) =\hat{P} \left[ \Phi^\mathrm{SC} \right] (l) $
 of exchange propagators. The matrix elements then read
 \begin{equation}\label{eqn:P_phi_sc}
 P_{m,n} (l) = \hat{P} \left[ \Phi^\mathrm{SC} \right]_{m,n} (l) =\left. \int \! d\mathbf{k} \, d\mathbf{k'} \, f^*_m(\mathbf{k}) \, f_n(\mathbf{k'}) \, \Phi^\mathrm{SC}_{l,k,k'} \right|_{k_0=k_0'=0} \, ,
\end{equation}
 and the single-channel coupling function is recovered as
\begin{equation} \label{eqn:app-PhiP}
 \Phi^\mathrm{SC}_{l,k,k'} \approx \sum_{m,n} f_m(\mathbf{k}) \, f^*_n(\mathbf{k'}) \, P_{m,n} (l) \, .
\end{equation}
 On a formal level, the momentum dependences are then reproduced exactly, while the frequency dependences contained in $k$ and $k'$ 
 are neglected which is expressed by the approximately-equal sign 
 in Eqn.~\ref{eqn:app-PhiP}.
 Additionally, in a numerical calculation, one will be forced to truncate the infinite sum over the form factors.
 Note that, if the underlying lattice structure corresponds to a Bravais lattice, the form factors can always be chosen to be real valued in momentum representation. 
 Hence, we will leave out the asterisks from Eqs.~\ref{eqn:P_phi_sc}~and~\ref{eqn:app-PhiP} in the following.

 Similarly to the particle-particle channel, the exchange propagators of the particle-hole channels are defined as
\begin{align}
    \label{eqn:C_phi_C}
 \mathbf{C} (l) &=\hat{C} \left[ \Phi^\mathrm{C} \right] (l)\, ,\\
    \label{eqn:D_phi_D}
 \mathbf{D} (l) &=\hat{D} \left[ \Phi^\mathrm{D} \right] (l)
\end{align}
 and the corresponding single-channel coupling functions read in exchange parametrization
\begin{align} \label{eqn:app-PhiC}
 \Phi^\mathrm{C}_{l,k,k'} \approx \sum_{m,n} f_m(\mathbf{k}) \, f_n(\mathbf{k'}) \, C_{m,n} (l) \, ,\\ \label{eqn:app-PhiD}
 \Phi^\mathrm{D}_{l,k,k'} \approx \sum_{m,n} f_m(\mathbf{k}) \, f_n(\mathbf{k'}) \, D_{m,n} (l) \, .
\end{align}

The flow equations for the exchange propagators are obtained by applying the projection operations from Eqs.~(\ref{eqn:P_phi_sc}), (\ref{eqn:C_phi_C}), 
and (\ref{eqn:D_phi_D}) to the respective diagrams in the right-hand sides of Eqs.~(\ref{eqn:flow-PhiP})-(\ref{eqn:flow-PhiD}). This yields
\begin{align} \label{eqn:expar-flowP}
 \dot{\mathbf{P}} (l) &= - \hat{P} \left[ \mathcal{T}_\mathrm{pp} \right] (l)\, ,\\ \label{eqn:expar-flowC}
 \dot{\mathbf{C}} (l) &=\hat{C} \left[ \mathcal{T}_\mathrm{ph}^\mathrm{cr} \right] (l)\, ,\\ \label{eqn:expar-flowD}
 \dot{\mathbf{D}} (l) &=\hat{D} \left[ \mathcal{T}_\mathrm{ph}^\mathrm{d} \right] (l)\, ,
\end{align}
 where the projection operators applied to a test function $F$ read:
\begin{align} \label{eqn:proj-P}
 \hat{P} \left[ F \right]_{m,n} (l) & = \left. \int \! d\mathbf{k} \, d\mathbf{k'} \,
f_m (\mathbf{k}) \,f_n (\mathbf{k'}) \, F \left(\frac{l}{2}+k,\frac{l}{2}-k,\frac{l}{2}-k'\right) \right|_{k_0=k_0'=0}  \, ,\\ \label{eqn:proj-C}
 \hat{C} \left[ F \right]_{m,n} (l) & = \left. \int \! d\mathbf{k} \, d\mathbf{k'} \,
f_m (\mathbf{k}) \,f_n (\mathbf{k'}) \, F \left(k+\frac{l}{2},k'-\frac{l}{2},k-\frac{l}{2}\right) \right|_{k_0=k_0'=0}  \, ,\\ \label{eqn:proj-D}
 \hat{D} \left[ F \right]_{m,n} (l) & = \left. \int \! d\mathbf{k} \, d\mathbf{k'} \,
f_m (\mathbf{k}) \,f_n (\mathbf{k'}) \, F \left(k+\frac{l}{2},k'-\frac{l}{2},k'+\frac{l}{2}\right) \right|_{k_0=k_0'=0}  \, .
\end{align}

\begin{figure}[t]
 \centering
 \includegraphics[width=.62\linewidth]{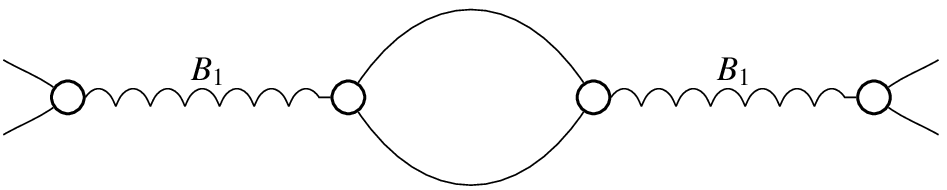}\\
 Propagator renormalization\\[3.9ex]
 \includegraphics[width=.55\linewidth]{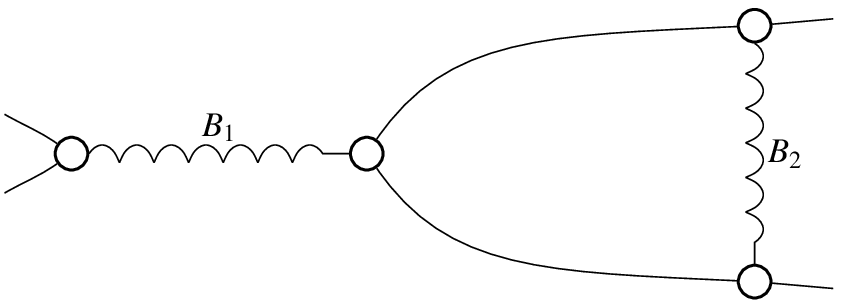}\\
 Vertex correction\\[3.9ex]
 \includegraphics[width=.28\linewidth]{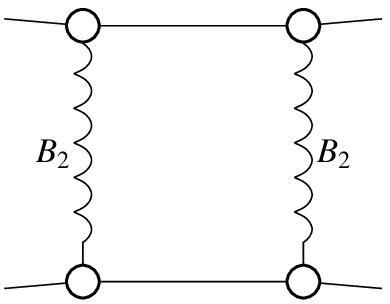}\\
 Box diagrams
 \caption{Diagrams contributing to the flow of $B_1$. Solid lines correspond to fermionic and wiggly ones to exchange propagators. Fermion-boson vertices
 (our form factors $f_m$) are represented by empty circles. $B_2$ denotes a linear combination of exchange propagators that differ from $B_1$.}
 \label{fig:expar-diags}
\end{figure}

Eqs.~(\ref{eqn:expar-flowP})-(\ref{eqn:expar-flowD}) can be seen as flow equations of the exchange-parametrization fRG
(see Refs.~\onlinecite{Husemann2009,Husemann2012,Giering2012} for details on this scheme).
By substituting the decomposed interaction of Eq.~(\ref{eqn:simple-deco}) in the flow equations (\ref{eqn:expar-flowP})-(\ref{eqn:expar-flowD}), one
obtains three different kinds of diagrams governing the flow of the exchange propagator $B_1 \in \{P,C,D\}$ (see Fig.~\ref{fig:expar-diags}). In the propagator renormalization
diagrams, bosonic lines (exchange propagators) corresponding to $B_1$ appear outside the loops.
Apart from a $B_1$ line outside the loops, vertex correction diagrams have one bosonic line inside, which corresponds to a linear combination $B_2$ of bosonic propagators. 
In the box diagrams, both bosonic lines correspond to $B_2$ and appear inside the loops. 
In a numerical implementation of the exchange-parametrization fRG, bosonic lines inside the loops pose a challenge. 
Since these exchange propagators may be sharply peaked close to an instability, they can significantly enhance the computational cost of the loop integrals.
In the following we show that one can separate the bosonic from the fermionic lines to make loop integrations a computationally easier task.

\subsection{Insertion of truncated partitions of unity}\label{sec:trunc_uni}
We continue with the derivation of the TUfRG equations by inserting partitions of unity of the form factor basis
\begin{equation}
    1=\int \! d\mathbf{p'} \, \delta(\mathbf{p}-\mathbf{p'})=\int \! d\mathbf{p'} \, \sum_m \, f_m(\mathbf{p'}) f_m(\mathbf{p})
    \label{eqn:unity}
\end{equation}
on both sides of the two fermion loops in Eqs.~(\ref{eqn:expar-flowP})-(\ref{eqn:expar-flowD}). These equations can be rewritten as
\begin{align} \label{eqn:matmul-P}
 \dot{\mathbf{P}} (l) &= \mathbf{V}^P (l)\, \dot{\boldsymbol\chi}^\mathrm{pp} (l) \, \mathbf{V}^P (l)\, ,\\ \label{eqn:matmul-C}
 \dot{\mathbf{C}} (l) &= - \mathbf{V}^C (l)\, \dot{\boldsymbol\chi}^\mathrm{ph} (l) \, \mathbf{V}^C (l)\, ,\\ \label{eqn:matmul-D}
 \dot{\mathbf{D}} (l) &=  2  \mathbf{V}^D (l)\, \dot{\boldsymbol\chi}^\mathrm{ph} (l) \, \mathbf{V}^D (l) -
   \mathbf{V}^C (l)\, \dot{\boldsymbol\chi}^\mathrm{ph} (l) \, \mathbf{V}^D (l) - \mathbf{V}^D (l)\, \dot{\boldsymbol\chi}^\mathrm{ph} (l) \, \mathbf{V}^C (l) \, ,
\end{align}
where 
\begin{equation}
\begin{split} \label{eqn:chis}
 \chi^\mathrm{pp}_{m,n} (l) &= \int \! dp \, G \left( \frac{l}{2} + p \right) \, G \left( \frac{l}{2} - p \right) \, f_m (\mathbf{p}) \, f_n (\mathbf{p}) \, ,\\
 \chi^\mathrm{ph}_{m,n} (l) &= \int \! dp \, G \left(p+ \frac{l}{2} \right) \, G \left(p- \frac{l}{2} \right) \, f_m (\mathbf{p}) \, f_n (\mathbf{p}) \, 
\end{split}
\end{equation}
and
\begin{align}
 \label{eqn:proj_VP}\mathbf{V}^P  \left(l \right) & = \hat{P} \left[ V^{(0)} \right] (l) - \mathbf{P} (l) + \hat{P} \left[ \Phi^\mathrm{C} \right] (l) + \hat{P} \left[ \Phi^\mathrm{D} \right] (l)  \, ,\\
 \label{eqn:proj_VC}\mathbf{V}^C  \left(l \right) & = \hat{C} \left[ V^{(0)} \right] (l) - \hat{C} \left[ \Phi^\mathrm{SC} \right] (l) + \mathbf{C} (l) + \hat{C} \left[ \Phi^\mathrm{D} \right] (l)  \, ,\\
 \label{eqn:proj_VD}\mathbf{V}^D  \left(l \right) & = \hat{D} \left[ V^{(0)} \right] (l) - \hat{D} \left[ \Phi^\mathrm{SC} \right] (l) + \hat{D} \left[ \Phi^\mathrm{C} \right] (l) + \mathbf{D} (l) 
\end{align}
are the three different projections from Eqs.~(\ref{eqn:proj-P})-(\ref{eqn:proj-D}) applied to the two-particle interaction.
Via Eqs.~(\ref{eqn:app-PhiP}), (\ref{eqn:app-PhiC}), and (\ref{eqn:app-PhiD}), the exchange propagators are inserted back into the flow equations~(\ref{eqn:matmul-P})-(\ref{eqn:matmul-D}), which
results in a closed system of differential equations. 
The bosonic propagators are now separated from the loop integrations at the cost of introducing the projections (\ref{eqn:proj_VP})-(\ref{eqn:proj_VD}).
For instance, the third contribution of $\mathbf{V}^P (l)$ can be expressed as 
\begin{align}
    \hat{P} \left[ \Phi^\mathrm{C} \right]_{m,n} (l) & \approx \int \! d\mathbf{k} \, d\mathbf{k'} \, f_m(\mathbf{k}) \, f_n(\mathbf{k'}) \notag \\ 
    & \qquad \times \, \left. \sum_{m',n'} \, f_{m'}\left(\frac{\mathbf{l}+\mathbf{k}-\mathbf{k'}}{2}\right) \, f_{n'}\left(\frac{\mathbf{l}-\mathbf{k}+\mathbf{k'}}{2}\right) \, 
    C_{m',n'}(k'+k) \right|_{k_0=k'_0=0}
    \label{eqn:proj_mom} \\
    & = \sum_{\mathbf{R_1},\mathbf{R_2},\mathbf{R_3}} \, \sum_{m',n'} \, f_m\left(-\frac{\mathbf{R_1}}{2}+\frac{\mathbf{R_2}}{2}-\mathbf{R_3}\right) \,
    f_n\left(\frac{\mathbf{R_1}}{2}-\frac{\mathbf{R_2}}{2}-\mathbf{R_3}\right) \notag \\
    & \qquad\qquad \times \, f_{m'}\left(\mathbf{R_1}\right) \, f_{n'}\left(\mathbf{R_2}\right) \, 
    C_{m',n'}(\mathbf{R_3},k_0=0) \, e^{-i\frac{1}{2}\mathbf{l}\cdot(\mathbf{R_1}+\mathbf{R_2})}
    \label{eqn:proj_pos}
\end{align}
in momentum and position space.

\begin{figure}[t]
 \centering
 \includegraphics[width=.35\linewidth]{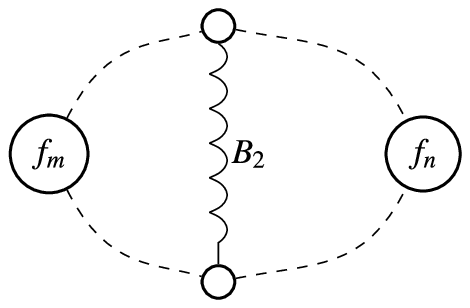}\\[1.9ex]
 {\Huge $\downarrow$}\\[1.9ex]
 \includegraphics[width=.76\linewidth]{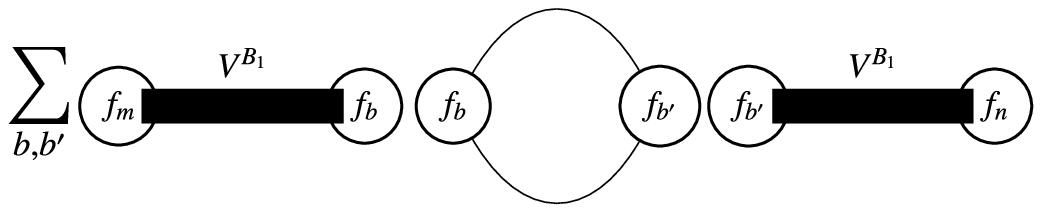}
 \caption{Graphical representation of the steps in the calculation of the increment for the interaction in the TUfRG approach. The upper diagram
 corresponds to the projection in step~\ref{enu:project-new}.) and the lower one to steps~\ref{enu:chis-new}.) and \ref{enu:matmul-new}.). The brick-stones in the lower part
    correspond to the \emph{full} interaction projected to the respective channel with $B_1= P$, $C$, or $D$. Dashed lines correspond to simple contractions and 
    \emph{not} to fermionic or bosonic propagators.}
 \label{fig:steps-new}
\end{figure}

Let us summarize which steps need to be performed in order to calculate the increment of the interaction in the TUfRG scheme:
\begin{enumerate}[i.)]
\item Project ${\mathbf{P}} (l)$, ${\mathbf{C}} (l)$, ${\mathbf{D}} (l)$ and the bare interaction to other channels in order to obtain 
     ${\mathbf{V}}^P (l)$, ${\mathbf{V}}^C (l)$, and ${\mathbf{V}}^D (l)$ according to Eqs.~(\ref{eqn:proj_VP})-(\ref{eqn:proj_VD}). 
    \label{enu:project-new}
\item Calculate the form factor projected fermionic loops $\dot{\boldsymbol\chi}^\mathrm{pp} (l)$ and $\dot{\boldsymbol\chi}^\mathrm{ph} (l)$ in Eq.~(\ref{eqn:chis}). \label{enu:chis-new}
\item Evaluate $\dot{\mathbf{P}} (l)$, $\dot{\mathbf{C}} (l)$, and $\dot{\mathbf{D}} (l)$ by performing matrix multiplications in the form factor basis 
    [see Eqs.~(\ref{eqn:matmul-P})-(\ref{eqn:matmul-D})]. 
     \label{enu:matmul-new}
\end{enumerate}
Fig.~\ref{fig:steps-new} graphically represents the calculation steps that are listed above.
 
 For implementing the TUfRG flow we require
\begin{enumerate}[a)]
 \item a suitable grid for the bosonic frequencies and momenta and
 \item a finite set of form factors $f_m (\mathbf{k})$.
\end{enumerate}
 While an inappropriate choice of the former may cause severe parametrization errors, the form factor expansion should be truncated in a way that avoids large
 projection errors. Generically, form factors corresponding to fixed bond lengths on the direct lattice seem appropriate, as it is likely and in fact known from previous studies (e.g.~\onlinecite{Maier2013}) that only small bond lengths (or slowly varying form factors) contribute significantly.

\subsection{Benefits from the truncated partitions of unity} \label{sec:benefits}
\begin{figure}[t]
\centering
\includegraphics[width=.84\linewidth]{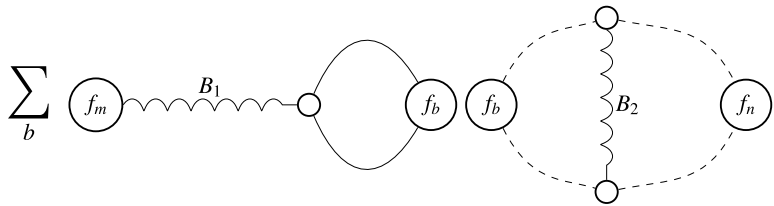}
\caption{Insertion of a unity operator in the form factor basis into a vertex-correction diagram.}
\label{fig:insert-one}
\end{figure}

From a formal point of view, our approach is nothing else than the standard exchange parametrization method with an additional approximation.
Namely, we have inserted truncated partitions of unity in the form factor basis in order to pull bosonic lines out of the loops.%
\footnote{Formally, our scheme reproduces the original one-loop flow equations of Sec.~\ref{sec:su2-inv} for a complete (infinite) set of form factors.}
This additional approximation is depicted in Fig.~\ref{fig:insert-one} for a vertex-correction diagram.
A suitable truncation of the form factor expansion is likely to contain more terms than in the exchange parametrization studies in
Refs.~\onlinecite{Husemann2009,Husemann2012,Giering2012}. However, we still expect fast convergence with increasing maximal bond length.
(For a more detailed discussion, see Section~\ref{sec:phase_hubb} and \ref{sec:app-smooth-ff}.)

The insertion of truncated unity partitions generates a computational advantage in calculating the right hand side of the flow equations due to the decomposition of the loop integrals.
As a consequence of separating the bosonic from the fermionic lines, the integrands only consist of a product of two fermionic propagators and two slowly varying form factors
instead of two interaction channels that can be strongly peaked close to a phase transition. Since integrations over structures with sharp features are numerically expensive, 
the replacement by smooth functions makes the loop integration an easier task. This simplification is done at the cost of adding the projection operations \ref{enu:project-new}.). 
As can be seen from Eq.~(\ref{eqn:proj_mom}), these additional tasks consist of two nested momentum integrals involving a product of 
four form factors and one exchange propagator. If the form factors correspond to fixed bond lengths, the calculation can be done most efficiently in position space~\cite{Wang2012}. In this 
case, the form factors translate to superpositions of Kronecker deltas that limit the appearing sums to a fixed upper bond length (see Eq.~(\ref{eqn:proj_pos})). 
This upper length is of course directly related to
the truncation length of the form factor basis. 

With these preliminary considerations, the projection step can be implemented in terms of evaluating overlaps of Kronecker-deltas and performing 
Fourier transforms of the exchange propagators for a finite set of vectors in position space. Moreover, it should be mentioned that the exchange propagator is the only object that 
depends on the fRG scale: all the other components (e.g. the non-vanishing Kronecker delta combinations and Fourier exponentials) stay the same for the whole fRG flow. 
Such a simplified dependence enables us to calculate these scale-independent parts only once and to use the result at all scales instead of repeating the same 
calculation at every step. We implemented the reuse of precomputed projection data in the code version for studies on the honeycomb lattice~\cite{Pena2016}.
In a calculation using $3217$ sampling points for each dependence on momentum $\mathbf{l}$ in Eqs.~(\ref{eqn:proj_VP})-(\ref{eqn:proj_VD}), the recycling of data caused speedups of 
$2.6$ and $1.7$ in the case of truncations after $7$ and $13$ form factors respectively. Although the total size of those data is of the order of some gigabytes and 
this part of the code is not yet optimized in terms of cache lines and load balancing, the 
computation time for the projections can be reduced by storing precomputed data.\footnote{See \ref{sec:symmetries} for details on how symmetries 
can be used for minimizing the memory consumption.} In case of a well behaved loop integrand, 
the projection process needs the major part of the computation time. Then, the storing of data can accelerate the whole fRG flow significantly. 
In terms of performance, the latter case might be seen as the optimal area of application for the TUfRG.

\section{Towards high-performance fRG}\label{sec:perform}
Despite the physically motivated truncations that enter the TUfRG, the
development of a computationally efficient method, that significantly
reduces the time-to-solution while providing meaningful predictions of
ground state properties, relies on the usage of high-performance
computers. In the last decade the evolution of the building blocks of
large computing architectures moved from single-core CPUs to compute nodes
with multiple cores, where large numbers of them are interconnected in
complex and heterogeneous networks.  As a consequence of
this evolution, it is only natural that a modern fRG implementation
should be able to make use of a large number of compute cores in order
to maximally exploit the parallelism of modern computing platforms.
To this
purpose, in our implementation, we make extensive use of the
directive-based OpenMP as well as the Message Passing Interface (MPI)
API, which are the most used standards for achieving shared memory and
distributed memory parallelization, respectively.

A clear advantage of the TUfRG method lies in the fact that bosonic
lines have been completely pulled out of the loops. Consequently, the
integrals in step~\ref{enu:chis-new}.)  are generically more well
behaved than in the exchange-parametrization approach (see
Section~\ref{sec:benefits}). Furthermore, all the loop integrations
for different form factor and bosonic momentum combinations are
completely independent from each other\footnote{Note that the same is
  true for the bosonic frequencies. However, in the current
  implementation we neglect the frequency dependence of the exchange
  propagators and focus on the zero frequency terms.}, and so
embarrassingly parallel. In this step,
communication is only necessary to share the results of the integration tasks between the
different MPI processes. When compared to the time spent in computations, the
communication overhead is negligible.
In our current implementation we use MPI for distributing the bosonic
momenta across the available compute nodes, i.e. the largest
computation unit whose constituents share one block of memory, and
OpenMP for parallelizing the form factor combinations of each bosonic
momentum.  To perform a single integration, for fixed form factor
indices and momentum, we use the adaptive quadrature routine
\texttt{DCUHRE}~\cite{Berntsen1991}.

For our implementation of step~\ref{enu:project-new}.) we used the
position space representation from Eq.~\eqref{eqn:proj_pos} in the
variant that avoids large memory consumptions.  Here we decided to
accept a longer runtime for this part of the calculation, since in the
$t$-$t'$ Hubbard model at van Hove filling step~\ref{enu:chis-new}.)
consumes the major part of the total computation time. As in
step~\ref{enu:chis-new}.) we use a hybrid parallelization, where the
sum over $m'$ and $n'$ from the right-hand side of
Eq.~\eqref{eqn:proj_pos} is distributed over the nodes with MPI, and the
different components of the exchange propagators regarding the form
factor indices $m$ and $n$ on the left-hand side are calculated in
parallel using OpenMP.

The matrix multiplications in \ref{enu:matmul-new}.) are of minor
relevance in terms of compute time and their implementation is 
therefore not optimized yet. It is based on nested \textit{for}-loops,
where those over external indices are parallelized using OpenMP.
For calculations using up to $128$ nodes, the compute time needed by this 
step is negligible compared to the ones of steps~\ref{enu:project-new}.) and 
\ref{enu:chis-new}.), due to the shared memory parallelism of the 
\textit{for}-loop iterations.

As explained above, by performing the steps~\ref{enu:project-new}.) -
\ref{enu:matmul-new}.) we calculate the derivative of the exchange
propagators with respect to the regularization scale. Since we aim to
obtain these propagators at lower scales, we are left with solving
ordinary differential equations (ODEs) of order one, which is a
standard task that we have implemented with the use of an explicit ODE
solver from the `Odeint' library~\cite{Ahnert2011}.

\begin{figure}[t]
    \centering
    \includegraphics[width=0.55\textwidth]{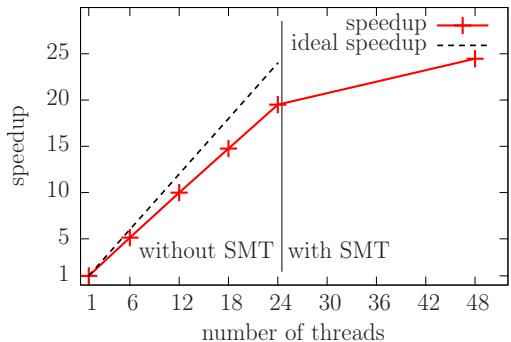}
    \caption{Speedup gained from shared memory parallelization relative to serial execution. The data have been produced using one node of a general purpose cluster with $24$ physical cores. For thread numbers higher than $24$ more than one thread is executed on one physical core.}
    \label{fig:speedup_threads}
\end{figure}
\begin{figure}[t]
    \centering
    \includegraphics[width=0.55\textwidth]{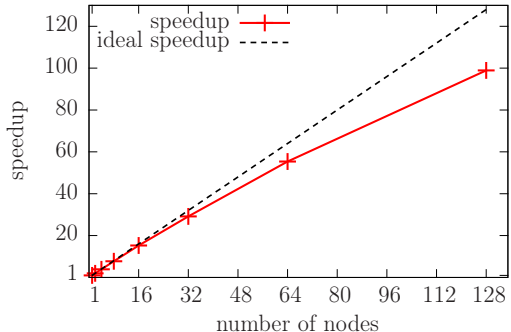}
    \caption{Speedup gained from distributed memory parallelization relative to an execution on one node. The data have been produced using a general purpose cluster with $24$ cores per node.}
    \label{fig:speedup_nodes}
\end{figure}

We analyzed our implementation of the TUfRG in terms of
parallelization speedup using the JURECA compute cluster, which 
is located at the J\"ulich Supercomputing Centre.
Every node consists of two Intel Xeon E5-2680 v3 Haswell CPUs with 
$12$ cores each, working at $2.5\, \textnormal{GHz}$. Due to 
simultaneous multithreading (SMT), every JURECA node supports 
$48$ threads in total. While there are nodes with larger main 
memory available within the cluster, we only used nodes with 
$128\,\textnormal{GB}$ as this appeared to be more than enough for our needs.
For the following tests, we apply the implementation to the 
$t$-$t'$~Hubbard model at van~Hove filling with
model parameters $t'=-0.3\,t$ and $U=3.0\,t$.  In this
first application of the scheme, we neglect the self-energy feedback
and use the $\Omega$-cutoff~\cite{Husemann2009} as regulator.  The form
factor basis is chosen in a way that every element corresponds to a
certain bond length\footnote{This still leaves some freedom for the
  specification of the form factors. In the current implementation
  these basis functions are chosen to transform according to the
  irreducible representations of the $C_{4v}$ point group, i.e., they
  fulfill $s$-, $p$-, $d$-, and $g$-symmetry, respectively.}  and the
truncation of that basis is introduced by an upper limit in the bond
length. More precisely, a truncation at the $n$th nearest neighbor
means that we only take form factors into account that correspond to
the $n$th nearest neighbor bonds or to shorter
bonds. Fig.~\ref{fig:speedup_threads} shows how the runtime for one
calculation of the ODEs' right-hand side decreases with increasing
number of threads running on one node, or in other words it shows the
performance of our shared memory parallelization. In this context the
quantity `speedup' can be understood as the quotient of the runtime
using the reference configuration, i.e. one thread, and the runtime
using the respective number of threads. For these data we used a
truncation of the form factor basis at the fourth nearest neighbor and
compared to the data from Section~\ref{sec:phase_hubb} we have reduced
the density of sampling points for bosonic momenta in order to get a
runtime below $24$ hours for the serial execution. As it can be seen
from Fig.~\ref{fig:speedup_threads}, the speedup scales well with the
number of threads in the region where the number of threads is less or
equal to $24$ and each thread runs exclusively on a physical core.
With $24$ threads and one thread per core we achieve a speedup of
$19.5$, while by
harvesting the additional underlying hardware parallelism when putting
two threads on a physical core we arrive at a speedup of $24.5$.

At the next level we can further enhance this result by using the
distributed memory parallelization. In Fig.~\ref{fig:speedup_nodes}
the speedup is plotted against the number of nodes, where the point of
reference is the runtime when using $48$ threads on one node. Since we
want to analyze the performance of our implementation under production
conditions, we now choose the same resolution of the bosonic momenta
as in Section~\ref{sec:phase_hubb} (see Fig.~\ref{fig:q-mesh}) and a
truncation at the fifth nearest neighbor. Due to huge consumptions of
time, these conditions have not been feasible for the program
executions for Fig.~\ref{fig:speedup_threads}. However, this does not
diminish the validity of our analysis, since an increase of the
resolution will lead to an enlarged number of parallelizable work
packages, which in turn rather supports parallelizability.
Our implementation scales very well up to $64$~nodes
and for $128$~nodes we still find a very reasonable speedup of $98.4$,
as shown in Fig.~\ref{fig:speedup_nodes}. 

\section{The $t$-$t'$ Hubbard model as a test case} \label{sec:phase_hubb}
\begin{figure}[t]
    \centering
    \includegraphics[width=0.42\textwidth]{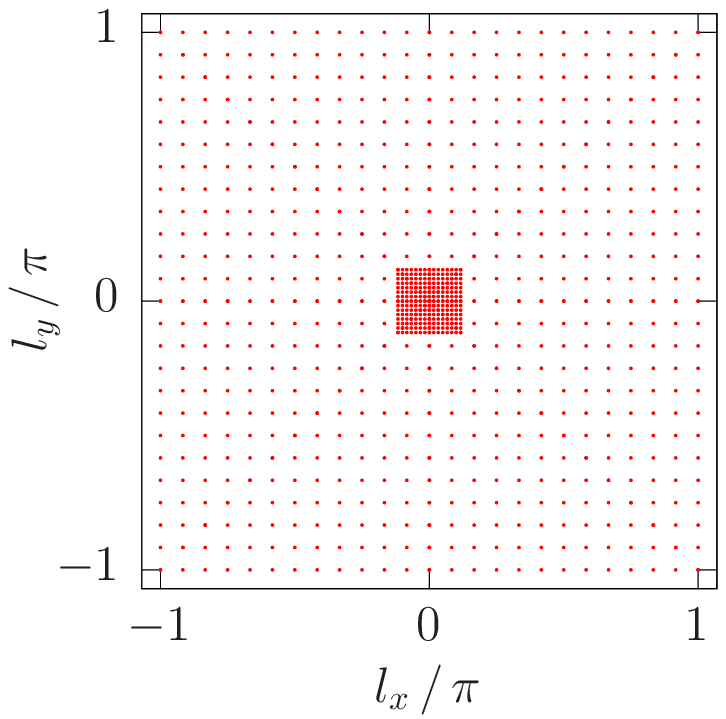}\hspace{2em}
    \includegraphics[width=0.42\textwidth]{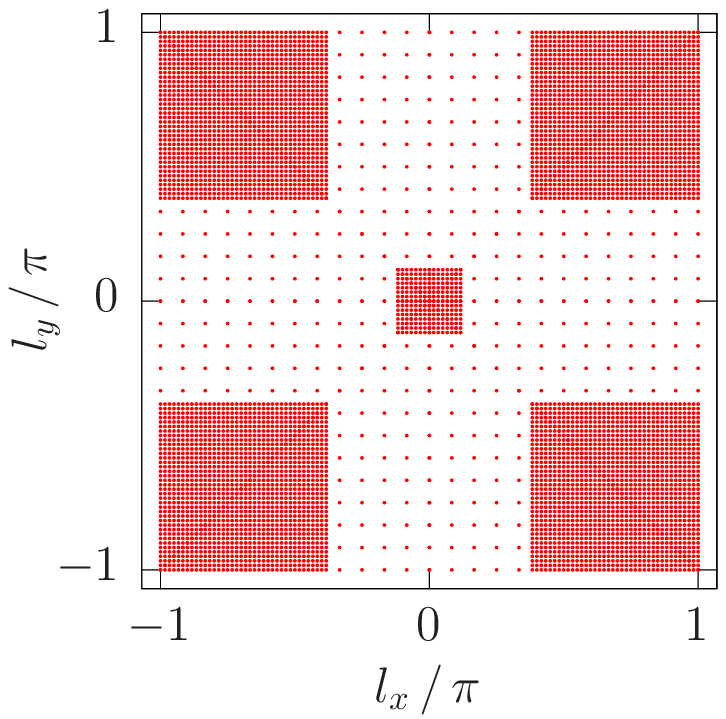}
    \caption{The discretization of the momentum dependence is shown for the particle-particle exchange propagator $\mathbf{P}(l)$ (left part, 992 grid points) and for the particle-hole exchange propagators $\mathbf{C}(l)$ and $\mathbf{D}(l)$ (right part, 6632 grid points).}
    \label{fig:q-mesh}
\end{figure}
We have applied our implementation of the TUfRG to the $t$-$t'$ Hubbard model on the square lattice, 
which is well studied~\cite{LeBlanc2015,Honerkamp2001,Giering2012,Eberlein2015,Yamase2016} but still contains some regions 
in the parameter space with unclear ground state properties. The single particle dispersion is given as
\begin{equation}
    \epsilon(\mathbf{k}) = -2\, t\, (\cos(k_x)+\cos(k_y)) - 4\, t'\, \cos(k_x)\, \cos(k_y) - \mu
    \label{eqn:hubb-disp}
\end{equation}
which contains three free parameters in general. In addition a fourth parameter in the Hubbard model is given by the onsite density-density interaction strength $U$. 
A simultaneous rescaling of these four parameters will leave the physics of the system invariant, but will rescale all physical energies. To take this 
into account we measure all energies relative to the parameter $t$ and leave the value of $t$ undefined. 
Furthermore, we restrict ourselves to van Hove filling $\mu = 4\,t'$ and use $U=3\,t$ 
which leaves us with only one free parameter $t'$.

Fig.~\ref{fig:q-mesh} shows the discretization of the momentum space that we have used for the calculation of the exchange propagators. 
Inside the areas of high grid point density we have expected strong peaks of the exchange propagator values that need to be resolved more accurately. 
Those areas have been chosen according to the results of previous studies on this model and can also be motivated by simple single channel deliberations.

\begin{figure}[t]
    \centering
    \includegraphics[width=0.85\textwidth]{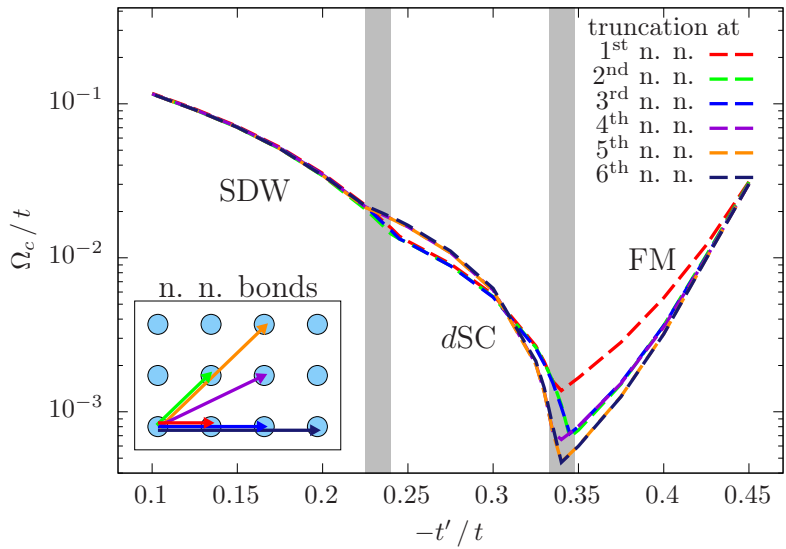}
    \caption{The critical scales for different truncation (bond) lengths of the form factor basis are plotted against the second nearest neighbor hopping $t'$ for the $t$-$t'$~Hubbard model at van~Hove filling with $U=3.0\,t$. Nearest neighbor bonds that correspond to the different truncations are shown in the inset. The gray bars separate the parameter regimes of the three observed instabilities, those are spin density wave (SDW) with $\mathbf{l}\sim(\pi,\pi)$, $d$-wave superconductivity ($d$SC) with zero total momentum and ferromagnetism (FM).}
    \label{fig:conv_test}
\end{figure}
In order to investigate the applicability of the insertion of truncated unity partitions, we have checked how the results change with increasing truncation length. 
To this end, we have performed the fRG flow in the 
parameter range from $t'=-0.10\,t$ to $t'=-0.45\,t$ with different truncations of the form factor basis. Starting from a truncation at the first nearest neighbor, 
we have successively increased the number of form factors until a sixth nearest neighbor truncation. 
The calculations have been started at an initial scale two orders of magnitude larger than the bandwidth and 
have been stopped when the largest absolute value of the exchange propagators has become one order of magnitude higher than the bandwidth.
It turns out that the results do not depend on the precise values of the initial scale and the largest exchange propagator component.
For each data point, the ground state type has been determined by the largest exchange propagator value by means of the corresponding channel (pairing, magnetic or charge), 
ordering vector, and form factor symmetry.
Fig.~\ref{fig:conv_test} shows the critical scales as functions of $t'$ and exhibits three different kinds of 
ground states. 
Spin density wave (SDW) and ferromagnetic (FM) instabilities manifest in the magnetic channel with an $s$-symmetry and ordering vector $\mathbf{l}\sim(\pi,\pi)$ 
and $\mathbf{l}=(0,0)$, respectively. The observed $d$-wave superconductivity ($d$SC) appears in the pairing channel with zero total momentum.
By reason of clarity the transitions between the different phases are only shown sketchily in this plot. 
However, the transition values of $t'$ turn out to change very mildly with the truncation length for both transitions. 

Focusing on the SDW regime, it becomes obvious that the critical scales are nearly unchanged by increasing the number of form factors. 
This shows that the important feedback from other channels---which lowers the critical scales compared to those from single channel calculations---%
is already contained in the TUfRG using a nearest neighbor truncation. Within the $d$SC regime we find a change of the critical scales 
by including the fourth nearest neighbor form factors. But this correction is small until we do not get too close to the phase transition to FM.
Moreover, it can be seen from Fig.~\ref{fig:conv_test} that numerically the last-named parameter region around $t'\approx-0.34\,t$ is the most difficult one. These 
difficulties are directly connected to the fact that the nature of this transition is highly unclear in general. 
There are fRG studies that find a quantum critical point between the two phases~\cite{Honerkamp2001,Giering2012} 
while others---like the present one---do not indicate such a phenomenon~\cite{Husemann2009}. By further decreasing $t'$ we enter the FM region, where a larger jump 
occurs between the scales of a first and a second nearest neighbor truncation. Beyond that, the inclusion of longer bond form factors has only a small impact on 
the results. Taken together we find a fast convergence of the critical scales---more precisely the influence of form factors beyond second nearest neighbor 
is rather low---within the investigated parameter regime except for values of $t'$ close to the phase transition between $d$SC and FM. 

Besides comparing different truncations with each other, a comparison to findings from other studies is necessary to validate the insertion of a 
truncated partition of unity into the flow equations. In Fig.~6 of Ref.~\onlinecite{Husemann2009} and in curve (i) in the left part of Fig.~3 of Ref.~\onlinecite{Husemann2012} 
the same model has been studied using exchange parametrization fRG in the same truncation of the flow equation hierarchy. The findings from those 
studies are very similar to our results in both the SDW and the $d$SC parameter regime, but in the region with a FM ground state those scales are closer to 
our results from first nearest neighbor truncation than to the higher order findings. Generally, it is not surprising that critical scales are reduced by increasing interchannel 
feedback---which is the consequence of taking more form factors into account---when the influence of the subleading ordering tendency ($d$SC) on the dominant one (FM) 
has destructive character.
Hence, our higher order truncation results can be seen as quantitative corrections to the critical scales from the two previous investigations within the FM region.

\section{Conclusion}
We have derived the TUfRG equations on the basis of an exchange parametrization in Sec.~\ref{sec:tufrg}.
As argued in Sec.~\ref{sec:benefits}, the loop integration in the TUfRG is much easier from a computational viewpoint 
than in a conventional exchange parametrization approach. This advantage has been obtained at the cost of adding a projection task which turns out to be 
of minor importance for the total computation time in many cases and can be accelerated by reusing precomputed data in the other cases. 
The convenience in numerics originates from a separation of fermionic and exchange propagators 
that at the same time simplifies the parallelization of the program code. Benefits from a hybrid parallelization in terms of speedup have 
been illustrated in Sec.~\ref{sec:perform}. As a consequence of accelerating the calculation by using many compute cores, 
it has been possible to access a large set of form factors in the TUfRG approach. 
Most importantly, a high-performance implementation combined with an efficient parametrization of the coupling function, as it is done in the TUfRG---originating from its 
relation to the exchange parametrization fRG---, should make applications to complex multiband systems possible.

A convergence check for the case of the $t$-$t'$ Hubbard model at van Hove filling has shown that the results converge fast with the number of form factors except for 
parameters close to the phase transition between $d$-wave superconductivity and ferromagnetism. In addition, we have seen a good agreement with results from previous exchange 
parametrization studies when using a comparable set of form factors. It has further been possible to take more form factors into account for producing results 
of higher accuracy.

\section*{Acknowledgments}
We thank Q.H. Wang and M. Salmhofer for discussions.
Numerical experiments have been conducted within the JUBE workflow environment~\cite{Luhrs2016} which has greatly facilitated data generation, management and analysis.
The authors gratefully acknowledge the computing time granted by JARA-HPC and provided on the supercomputer JURECA at J\"ulich Supercomputing Centre (JSC). Furthermore, the German Research Foundation (DFG) is acknowledged for support via RTG 1995 and SPP 1459.

\begin{appendix}

\section{Relation to the SMFRG} \label{sec:rel_to_smfrg}
In the following we describe the relation of the TUfRG equations to the ones used in the SMFRG, which was introduced in Ref.~\onlinecite{Wang2012}.
 Instead of exchange propagators, the core objects now are three complementary approximate representations of the two-particle interaction $V (k_1,k_2,k_3)$:
\begin{align*}
 V \left(k_1,k_2,k_3 \right) &\approx \sum_{m,n} f_m \left(\frac{\mathbf{k}_1-\mathbf{k}_2}{2} \right) \,f_n \left(\frac{\mathbf{k}_4-\mathbf{k}_3}{2} \right) \, V^P_{m,n}  \left(k_1+k_2 \right) \, ,\\
 V \left(k_1,k_2,k_3 \right) &\approx \sum_{m,n} f_m \left(\frac{\mathbf{k}_1+\mathbf{k}_3}{2} \right) \,f_n \left(\frac{\mathbf{k}_2+\mathbf{k}_4}{2} \right) \, V^C_{m,n}  \left(k_1-k_3 \right) \, ,\\
 V \left(k_1,k_2,k_3 \right) &\approx \sum_{m,n} f_m \left(\frac{\mathbf{k}_1+\mathbf{k}_4}{2} \right) \,f_n \left(\frac{\mathbf{k}_2+\mathbf{k}_3}{2} \right) \, V^D_{m,n}  \left(k_3-k_2 \right) \, .
\end{align*}
 The form of these three representations is reminiscent of the exchange parametrization of the single-channel coupling functions in
 Eqs.~(\ref{eqn:app-PhiP})-(\ref{eqn:app-PhiD}).
 However, we are dealing with three different approximations of the same coupling function in the present case, and it depends on the context which one is used.

 Formally, the bond length representations
\begin{align*}
 \mathbf{V}^P  \left(k_1+k_2 \right) & = \hat{P} \left[ V \right] (k_1 + k_2) \, ,\\
 \mathbf{V}^C  \left(k_1-k_3 \right) & = \hat{C} \left[ V \right] (k_1 - k_3) \, ,\\
 \mathbf{V}^D  \left(k_3-k_2 \right) & = \hat{D} \left[ V \right] (k_3 - k_2)
\end{align*}
 can be projected out from $V$ by applying the projection rules (\ref{eqn:proj-P})-(\ref{eqn:proj-D}). From these relations one can directly see that those are the same objects 
 as the ones calculated in Eqs.~(\ref{eqn:proj_VP})-(\ref{eqn:proj_VD}) as part of the TUfRG procedure.
 In contrast to Refs.~\onlinecite{Wang2012,Wang2013,Xiang2013}, the conventions used here render the matrices $\mathbf{V}^P$, $\mathbf{V}^C$, and $\mathbf{V}^D$ 
 hermitian due to the Pauli principle and positivity.
 Their flow arises from a projection of all five one-loop diagrams of Fig.~\ref{fig:diags-fleq} to the respective channel:
\begin{align*}
 \dot{\mathbf{V}}^P  \left(l \right) & = \hat{P} \left[ \mathcal{T}_\mathrm{pp} + \mathcal{T}_\mathrm{ph}^\mathrm{cr} + \mathcal{T}_\mathrm{ph}^\mathrm{d} \right] (l) \, ,\\
 \dot{\mathbf{V}}^C  \left(l \right) & = \hat{C} \left[ \mathcal{T}_\mathrm{pp} + \mathcal{T}_\mathrm{ph}^\mathrm{cr} + \mathcal{T}_\mathrm{ph}^\mathrm{d} \right] (l) \, ,\\
 \dot{\mathbf{V}}^D  \left(l \right) & = \hat{D} \left[ \mathcal{T}_\mathrm{pp} + \mathcal{T}_\mathrm{ph}^\mathrm{cr} + \mathcal{T}_\mathrm{ph}^\mathrm{d} \right] (l) \, .
\end{align*}
This is nothing but the derivatives of Eqs.~(\ref{eqn:proj_VP})-(\ref{eqn:proj_VD}) with respect to the regularization scale:
\begin{align}
    \label{equ:VP_dot}
    \dot{\mathbf{V}}^P  \left(l \right) & = - \dot{\mathbf{P}} (l) + \hat{P} \left[ \dot{\Phi}^\mathrm{C} \right] (l) + \hat{P} \left[ \dot{\Phi}^\mathrm{D} \right] (l)  \, ,\\
    \label{equ:VC_dot}
    \dot{\mathbf{V}}^C  \left(l \right) & = - \hat{C} \left[ \dot{\Phi}^\mathrm{SC} \right] (l) + \dot{\mathbf{C}} (l) + \hat{C} \left[ \dot{\Phi}^\mathrm{D} \right] (l)  \, ,\\
    \label{equ:VD_dot}
    \dot{\mathbf{V}}^D  \left(l \right) & = - \hat{D} \left[ \dot{\Phi}^\mathrm{SC} \right] (l) + \hat{D} \left[ \dot{\Phi}^\mathrm{C} \right] (l) + \dot{\mathbf{D}} (l) \, .
\end{align}
Scale derivatives of single channel coupling functions $\Phi^\mathrm{SC}$, $\Phi^\mathrm{C}$ and $\Phi^\mathrm{D}$ can be expressed 
by derivatives of exchange propagators as implied by Eqs.~(\ref{eqn:app-PhiP})-(\ref{eqn:app-PhiD}). Using Eqs.~(\ref{eqn:matmul-P})-(\ref{eqn:matmul-D}) we obtain a 
closed system of differential equations for $\mathbf{V}^P$, $\mathbf{V}^C$, and $\mathbf{V}^D$. While in the TUfRG the exchange propagators are the central objects that are 
stored during the whole fRG calculation, in the SMFRG only the derivatives can be known. Those values have to be calculated in every SMFRG step, since they are 
needed temporarily for calculating the right-hand sides of Eqs.~(\ref{equ:VP_dot})-(\ref{equ:VD_dot}).
\begin{figure}[t]
    \centering
    \includegraphics[width=.76\linewidth]{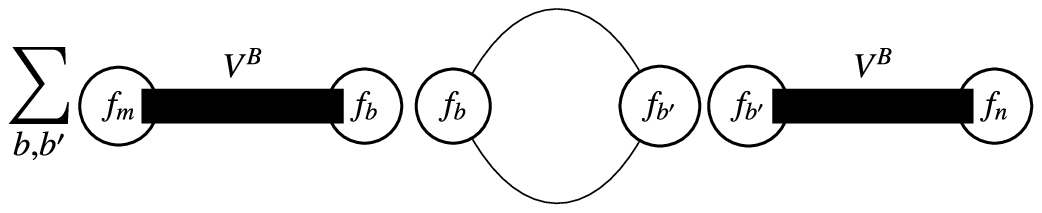}\\[1.9ex]
    {\Huge $\downarrow$}\\[1.9ex]
    \includegraphics[width=.35\linewidth]{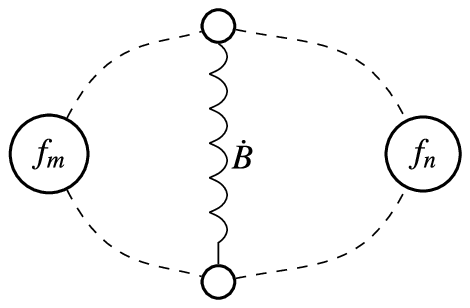}
    \caption{Graphical representation of the steps in the calculation of the increment for the interaction in the SMFRG approach. The upper diagram
    corresponds to steps~\ref{enu:chis}.) and \ref{enu:matmul}.) and the lower one to the projection in step~\ref{enu:project}.). The brick-stones in the upper part
    correspond to the \emph{full} interaction projected to the respective channel with $B= P$, $C$, or $D$. Dashed lines correspond to simple contractions and 
    \emph{not} to fermionic or bosonic propagators.}
    \label{fig:steps-smfrg}
\end{figure}
In summary, calculating the increment in the SMFRG flow of the interaction involves the following steps
(for a graphical representation, see Fig.~\ref{fig:steps-smfrg}):
\begin{enumerate}[1.)]
    \item Calculate the form factor projected fermionic loops $\dot{\boldsymbol\chi}^\mathrm{pp} (l)$ and $\dot{\boldsymbol\chi}^\mathrm{ph} (l)$ in Eq.~(\ref{eqn:chis}). \label{enu:chis}
    \item Evaluate $\dot{\mathbf{P}} (l)$, $\dot{\mathbf{C}} (l)$, and $\dot{\mathbf{D}} (l)$ by performing matrix multiplications in the form factor basis [see Eqs.~(\ref{eqn:matmul-P})-(\ref{eqn:matmul-D})]. \label{enu:matmul}
    \item Project $\dot{\mathbf{P}} (l)$, $\dot{\mathbf{C}} (l)$, and $\dot{\mathbf{D}} (l)$ to other channels and obtain $\dot{\mathbf{V}}^P (l)$, $\dot{\mathbf{V}}^C (l)$, and $\dot{\mathbf{V}}^D (l)$ according to Eqs.~(\ref{equ:VP_dot})-(\ref{equ:VD_dot}). \label{enu:project}
\end{enumerate}
From a computational viewpoint these steps are the same as the ones from Sec.~\ref{sec:trunc_uni}, just in a different order of execution. Therefore, the computational effort is the 
same for both schemes.

Taken together, the main difference between TUfRG and SMFRG is the choice of the core objects. Within the TUfRG framework we permanently store the exchange propagators during 
the flow, as it is done in exchange parametrization studies like~\onlinecite{Husemann2009,Husemann2012,Giering2012}. In contrast, the three 
different projections of $V$ play this role within the SMFRG. From Eqs.~(\ref{eqn:proj_VP})-(\ref{eqn:proj_VD}) we can directly obtain the projected $V$s from the exchange propagators 
in the TUfRG scheme and thus should be able to recover the SMFRG results. Vice versa, it would be necessary to invert those equations in order 
to extract the TUfRG results from the SMFRG. This is nothing but a matrix inversion, which is a very expensive task, since the dimension of that matrix 
is equal to the total number of components of the projected couplings or of the exchange propagators, respectively. 
E.g., in the case of a truncation at the fifth nearest neighbor and the same momentum resolution as in Section~\ref{sec:phase_hubb} one would need to 
invert a matrix with a dimension of about $8.9 \times 10^6$. 

Let us now compare the applicability of the projected couplings and the 
exchange propagators in the context of recovering the full coupling function, which is needed for self-energy calculations, e.g.. On the side of the TUfRG there is 
a unique way of calculating the single-channel coupling functions from the exchange propagators and from those the full coupling. The three strong momentum dependences, 
one from each channel, directly enter the resulting object, since it is just a sum of the three channels. On the side of the SMFRG there are three different 
possibilities to undo the projection, i.e. we can use $\mathbf{V}^P$, $\mathbf{V}^C$, or $\mathbf{V}^D$ to get $V$. The result will strongly depend on the choice we have used for 
the calculation, since the projected couplings only contain one of the three important momentum dependencies while the other two are smoothened out by the 
projection process and can not be recovered due to a truncated form factor basis.

With the last two paragraphs in mind and in view of the fact that the numerical effort is the same, we see a clear advantage in using the exchange propagators 
as central objects instead of the projected couplings.

\section{Applicability of the truncation of form factor unity operators} \label{sec:app-smooth-ff}

 In the following, an argument for the applicability of the insertion of truncated unity partitions in vertex-correction and box diagrams is given.
 
 For this purpose, let us rewrite the rule~(\ref{eqn:proj-D}) for the projection to the direct particle-hole channel in the following way:
\begin{equation}  \label{eqn:internal-proj}
 \hat{D} \left[ F \right]_{m,n} (l) = \left. \int \! d\mathbf{k'} \, f_n (\mathbf{k'}) \, \, \hat{E}_D \left[F \right]_m (l,k') \right|_{k'_0=0} \, ,
\end{equation}
 where the external projections
 \begin{equation}\label{eqn:external-proj}
 \hat{E}_D \left[ F \right]_m (l,k') = \left. \int \! d\mathbf{k} \,
f_m (\mathbf{k}) \, F \left(k+\frac{l}{2},k'-\frac{l}{2},k'+\frac{l}{2}\right) \right|_{k_0=0}
\end{equation}
 have been defined in a similar way as in Ref.~\onlinecite{Giering2012}.
 While the contractions of the coupling function $F$ with the two form factors $f_m (\mathbf{k})$ and $f_n (\mathbf{k'})$ have been treated on equal footing in Eq.~(\ref{eqn:proj-D}),
 the contraction with the form factor $f_n (\mathbf{k'})$ resulting from the insertion of a truncated partition of unity is
 performed \emph{after} the external projection $\hat{E}_D [F]_m (l,k') $ in Eq.~(\ref{eqn:internal-proj}).
 In the following, we will argue that, for external form factors corresponding to short bond lengths,
 the external projections $ \hat{E}_D [F]_m (l,k') $ vary slowly in $\mathbf{k}'$. The sum over the inserted form factors $f_n (\mathbf{k'})$ can therefore
 be truncated after a certain bond length, since contributions from longer bonds vanish.

 Here, we focus on the feedback of the particle-particle and the crossed particle-hole channels on the direct particle-hole channel.
 We therefore consider Eq.~(\ref{eqn:external-proj}) with 
\begin{align*}
 F (k_1,k_2,k_3) &= \sum_{c,c'} f_c \left( \frac{\mathbf{k}_1 - \mathbf{k}_2}{2} \right) \, f_{c'} \left( \frac{\mathbf{k}_4 - \mathbf{k}_3}{2} \right) \, P_{c,c'} \left( k_1 + k_2 \right) \\
 &  \left. \quad + \sum_{c,c'} f_c \left( \frac{\mathbf{k}_1 + \mathbf{k}_3}{2} \right) \, f_{c'} \left( \frac{\mathbf{k}_2 + \mathbf{k}_4}{2} \right) \, C_{c,c'} \left( k_1 - k_3 \right) \right|_{k_4=k_1+k_2-k_3} \, .
\end{align*}
 However, this example is generic for the feedback between different interaction channels.
 For the external projection, we then have
\begin{align*}
 \hat{E}_D \left[ F \right]_m (l,k') &= \int \! d\mathbf{k} \, f_m (\mathbf{k}) \left[
 \sum_{c,c'} f_c \left( \frac{\mathbf{k} - \mathbf{k'} +\mathbf{l}}{2} \right) \, f_{c'} \left( \frac{\mathbf{k} - \mathbf{k'} - \mathbf{l}}{2} \right) \, P_{c,c'} \left( k + k' \right) \right. \\
 &  \left. \qquad + \sum_{c,c'} f_c \left( \frac{\mathbf{k} + \mathbf{k'} + \mathbf{l}}{2} \right) \, f_{c'} \left( \frac{\mathbf{k} + \mathbf{k'} - \mathbf{l}}{2} \right) \, C_{c,c'} \left( k - k' \right) \right]_{k_0=0} \, .
\end{align*}
 Apparently, if the exchange propagators $P$ and $C$ are slowly varying functions of the total and transfer momenta,
 the dependence of $\hat{E}_D \left[ F \right]_m (l,k')$ on $\mathbf{k'}$ is smooth.
 But also if these exchange propagators are strongly peaked, the strong dependence on $\mathbf{k'}$ is washed out by the above convolution-like integral.
 This can be more clearly seen after a substitution of the integration variable:
 \begin{align*}
 \hat{E}_D \left[ F \right]_m (l,k') &= \int \! d\mathbf{u} \, \left. f_m (\mathbf{u} - \mathbf{k'})
 \sum_{c,c'} f_c \left( \frac{\mathbf{u} - 2\mathbf{k'} +\mathbf{l}}{2} \right) \, f_{c'} \left( \frac{\mathbf{u} - 2\mathbf{k'} - \mathbf{l}}{2} \right) \, P_{c,c'} \left( u \right) \right|_{u_0=k'_0} \\
 &+ \int \! d\mathbf{u} \, \left. f_m (\mathbf{u} + \mathbf{k'})
\sum_{c,c'} f_c \left( \frac{\mathbf{u} + 2\mathbf{k'} + \mathbf{l}}{2} \right) \, f_{c'} \left( \frac{\mathbf{u} + 2\mathbf{k'} - \mathbf{l}}{2} \right) \, C_{c,c'} \left( u \right) \right|_{u_0=-k'_0} \, .
\end{align*}
 Let us now consider form factors
\begin{equation*}
 f_r (\mathbf{q}) = \sum_\mathbf{R} f_r (\mathbf{R}) \, e^{-i \mathbf{q} \cdot \mathbf{R}} 
\end{equation*}
 corresponding to fixed bond lengths. This means that $f_r (\mathbf{R}) $ only contributes for a fixed value of $| \mathbf{R} |$,
 which is given by $r$.
 We then can easily observe that the external projection $\hat{E}_D \left[ F \right]_m (l,k')$ varies smoothly with $\mathbf{k'}$.
 More formally, in the internal projection $\hat{D} [F]_{m,n} (l)$ in Eq.~(\ref{eqn:internal-proj}),
 contributions with values of $n$ that correspond to a large bond length will be absent.
 In particular, the projection integral vanishes once the bond length of $f_n (\mathbf{k'})$ exceeds
 the sum of the maximal bond lengths of the form factors labeled with $m$, $c$, and $c'$.

 This argument can be straightforwardly carried over to all other inter-channel feedback contributions in the vertex-correction and box diagrams in the fRG flow equations.
 Therefore, a numerically tractable truncation of the unity operators inserted between internal fermionic and bosonic lines should be applicable.

\section{Useful symmetries} \label{sec:symmetries}
 The spin-independent coupling function $V(k_1,k_2,k_3)$ obeys the relation
\begin{equation*}
 V( k_1,k_2,k_3) = V(k_2,k_1,k_1+k_2-k_3)
\end{equation*}
 stemming from the Pauli principle and the positivity of the action corresponding to the Hermiticity of the Hamiltonian requires
\begin{equation*}
 V( k_1,k_2,k_3) = V(\hat{k}_1+\hat{k}_2-\hat{k}_3,\hat{k}_3,\hat{k}_2)^\ast \, ,
\end{equation*}
where $ \hat{k} = (-k_0,\mathbf{k}) $ (see Appendix~A of Ref.~\onlinecite{Eberlein2014a} for a discussion). In the channel decomposed form this results in
\begin{align*}
 \Phi^\mathrm{SC}_{l,q,q'} &= \Phi^\mathrm{SC}_{l,-q,-q'} = \Phi^\mathrm{SC}_{l,q',q}\, ,\\
 \Phi^\mathrm{C}_{l,q,q'} &= \Phi^\mathrm{C}_{-l,q',q} = \Phi^\mathrm{C}_{l,q',q}\, ,\\
 \Phi^\mathrm{D}_{l,q,q'} &= \Phi^\mathrm{D}_{-l,q',q} = \Phi^\mathrm{D}_{-l,q,q'} \, ,
\end{align*}
 where the first equality sign in each line corresponds to the Pauli principle and the second one to positivity.
 Note that, for the latter, we have already exploited that all coupling functions are mapped to their complex conjugates under frequency inversion.
Using these relations with regard to exchange propagators it can be shown 
that the matrices $\mathbf{P}$, $\mathbf{C}$, and $\mathbf{D}$ are symmetric (cf.~Ref.~\onlinecite{Husemann2009}):
\begin{equation*}
 B_{m,n} (l) = B_{n,m} (l) \, , \quad \text{where} \quad B \in \{ P,C,D\} \, .
\end{equation*}
In addition, the constraints
\begin{align*}
 P_{m,n} (l) &= \pi_m \pi_n \, P_{m,n} (l) \, , \\
 C_{m,n} (-l) &= C_{m,n} (l) \, ,\\
 D_{m,n} (-l) &= D_{m,n} (l)
\end{align*}
hold, where $\pi_m$ denotes the parity eigenvalue of the $m$-th form factor, i.e., $ f_m (- \mathbf{k}) = \pi_m \, f_m (\mathbf{k})$.
 If the form factors are chosen to transform with irreducible representations of the point group, some matrix elements of $\mathbf{P}$, $\mathbf{C}$, and $\mathbf{D}$ vanish
 at points of high symmetry due to Schur's second lemma.~\cite{Maier2013}
 Under frequency inversion, all three exchange propagators behave as
\begin{equation*}
 B_{m,n} (\hat{l}) =B_{m,n} (l)^\ast \, , \quad \text{where} \quad B \in \{ P,C,D\} \, . 
\end{equation*}

For the aforementioned possibility of calculating the scale independent parts of the projection operations at the start of the flow and storing them, 
exploiting symmetries can reduce the substantial increase in memory usage. Such projection operations can be expressed as
\begin{equation*}
\hat{B}\left[ \Phi^\mathrm{B'} \right]_{m,n}(l)=\sum_{\substack{m',n' \\ \mathbf{q}}} A_{m,n,m',n'}^{B,B'}(\mathbf{q}, \mathbf{l}) B'_{m',n'}(\mathbf{q},q_0=0)\, ,
\end{equation*}
where---compared to Eqn.~\ref{eqn:proj_pos}---we have replaced the exchange propagator in position space by its Fourier series.
$\mathbf{A}^{B,B'}$ is constant during the flow, and corresponds to 
\begin{align*}
    A_{m,n,m',n'}^{P,C}(\mathbf{q}, \mathbf{l}) = 
    \sum_{\mathbf{R_1},\mathbf{R_2},\mathbf{R_3}} \, &f_m\left(-\frac{\mathbf{R_1}}{2}+\frac{\mathbf{R_2}}{2}-\mathbf{R_3}\right) \,
    f_n\left(\frac{\mathbf{R_1}}{2}-\frac{\mathbf{R_2}}{2}-\mathbf{R_3}\right) \,\\
    \times \  &f_{m'}\left(\mathbf{R_1}\right) \, f_{n'}\left(\mathbf{R_2}\right) \, 
    e^{i \mathbf{q} \mathbf{R_3}} \, e^{-i\frac{1}{2}\mathbf{l}\cdot(\mathbf{R_1}+\mathbf{R_2})}
\end{align*}
in the example case of feedback from the $C$ channel to the $P$ channel. The objects $\mathbf{A}^{B,B'}$ are all real valued, 
and the 6 possible combinations of $B,B'$ are related in pairs
\begin{align*}
\mathbf{A}^{P,C}&=\pi_n \mathbf{A}^{P,D}\, ,\\
\mathbf{A}^{C,P}&=\pi_{n'} \mathbf{A}^{D,P}\, ,\\
\mathbf{A}^{D,C}&= \mathbf{A}^{C,D}
\end{align*}
so that only three of them have to be computed and stored. Furthermore, each of these three objects has given symmetries with respect to exchange of form factor indices 
or momentum inversion. For the previous example those symmetries can be written as 
\begin{align*}
A_{n,m,m',n'}^{P,C}(\mathbf{q}, \mathbf{l})&= \pi_m \pi_n A_{m,n,m',n'}^{P,C}(\mathbf{-q}, \mathbf{l})\, ,\\
A_{m,n,n',m'}^{P,C}(\mathbf{q}, \mathbf{l})&= \pi_m \pi_n A_{m,n,m',n'}^{P,C}(\mathbf{-q}, \mathbf{l})\, ,\\
A_{m,n,m',n'}^{P,C}(\mathbf{-q}, \mathbf{-l})&= \pi_m \pi_n \pi_{m'} \pi_{n'} A_{m,n,m',n'}^{P,C}(\mathbf{q}, \mathbf{l})\, .
\end{align*}
The last-named equation can further be generalized by using the point group's symmetry operations. In a first step for each operation $\hat S$ a matrix has to be set up
that contains prefactors of the linear combination of form factors necessary to compose $f_m\left(\hat S \, \mathbf k \right)$, i.e.,
\begin{equation*}
    f_m\left(\hat S\, \mathbf k \right) \, = \, \sum_{\tilde m} \, C^{\hat S}_{m,\tilde m} \, f_{\tilde m}\left(\mathbf k \right) \, .
\end{equation*}
As a second step we can relate the elements of $\mathbf{A}^{B,B'}$ to the symmetry transformed ones, e.g.,
\begin{equation*}
    A_{m,n,m',n'}^{P,C}(\hat S \, \mathbf{q}, \hat S \, \mathbf{l})= \sum_{\tilde m,\tilde n,\tilde m',\tilde n'} C^{\hat S}_{m,\tilde m} \, C^{\hat S}_{n,\tilde n} \, C^{\hat S}_{m',\tilde m'} \, C^{\hat S}_{n',\tilde n'} \, A_{\tilde m,\tilde n,\tilde m',\tilde n'}^{P,C}(\mathbf{q}, \mathbf{l})\, .
\end{equation*}

The definitions of the form factor projected loops from Eqn.~\ref{eqn:chis} directly indicate that the matrices $\boldsymbol{\chi}^\mathrm{pp}$ and $\boldsymbol{\chi}^\mathrm{ph}$ are symmetric
\begin{align*}
    \chi^\mathrm{pp}_{n,m}(l) \, &= \, \chi^\mathrm{pp}_{m,n}(l) \, , \\
    \chi^\mathrm{ph}_{n,m}(l) \, &= \, \chi^\mathrm{ph}_{m,n}(l) \, ,
\end{align*}
and an inversion of the integration variable shows that in the particle-particle case only elements corresponding to form factors with the same parity may be nonzero 
\begin{equation*}
    \chi^\mathrm{pp}_{m,n}(l) \, = \, \pi_m \, \pi_n \, \chi^\mathrm{pp}_{m,n}(l) \, .
\end{equation*}
The usage of point group symmetries yields 
\begin{equation*}
    \chi^\mathrm{pp/ph}_{m,n} \left( \hat S \, \mathbf{l},l_0 \right) \, = \, \sum_{\tilde m,\tilde n} \, C^{\hat S}_{m,\tilde m} \, C^{\hat S}_{n,\tilde n} \, \chi^\mathrm{pp/ph}_{\tilde m,\tilde n} \left( \mathbf{l},l_0 \right) \, .
\end{equation*}

\end{appendix}

\bibliographystyle{apsrev4-1}
\bibliography{bibliography}
\end{document}